\documentclass[preprint, showpacs, superscriptaddress, aps]{revtex4-1}

\usepackage{amsmath}
\usepackage{graphicx}
\usepackage{amssymb}
\usepackage{graphicx}
\usepackage{subfigure}
\usepackage{hyperref}
\usepackage{verbatim}
 
\newcommand{\lb}{{<}}
\newcommand{\rb}{{>}}

\begin{document}

\title{Novel effective ergodicity breaking phase transition in a driven-dissipative system}
\author{Sakib Matin }

\affiliation{Department of Physics, Boston University, Boston, Massachusetts 02215, USA}
\author{Chon-Kit Pun}
\affiliation{Department of Physics, Boston University, Boston, Massachusetts 02215, USA}

\author{Harvey Gould}
\affiliation{Department of Physics, Boston University, Boston, Massachusetts 02215, USA}
\affiliation{Department of Physics, Clark University, Worcester, Massachusetts 01610, USA}

\author{W. Klein}
\affiliation{Department of Physics, Boston University, Boston, Massachusetts 02215, USA}
\affiliation{Center for Computational Science, Boston
University, Boston, Massachusetts 02215, USA}

\date{\today}

\begin{abstract}
We show that the Olami-Feder-Christensen model exhibits an effective ergodicity breaking transition as the noise is varied. Above the critical noise, the system is effectively ergodic because the time-averaged stress on each site converges to the global spatial average. In contrast, below the critical noise, the stress on individual sites becomes trapped in different limit cycles, and the system is not ergodic. To characterize this transition, we use ideas from the study of dynamical systems and compute recurrence plots and the recurrence rate. The order parameter is identified as the recurrence rate averaged over all sites and exhibits a jump at the critical noise. We also use ideas from percolation theory and analyze the clusters of failed sites to find numerical evidence that the transition, when approached from above, can be characterized by exponents that are consistent with hyperscaling.
\end{abstract}

\maketitle

\section{Introduction\label{sec:intro}}
The Fermi-Pasta-Ulam-Tsingou model~\cite{FPU_Review} is a well known example of a system that exhibits broken ergodicity. The model exhibits quasi-regular dynamics below the critical energy threshold and reaches equipartition above the threshold. Understanding the nature of this type of transition may help extend the tools of equilibrium statistical mechanics to driven-dissipative systems~\cite{Olami_1}, active matter~\cite{vicsek}, and other nonequilibrium phenomena.

In this paper we consider the nearest-neighbor Olami-Feder-Christensen (OFC) model~\cite{Olami_1}. This model is a driven dissipative system that has been of particular interest in the context of the study of earthquakes. We simulate the OFC model on a square lattice of length $L$ with $N=L^2$ sites and periodic boundary conditions. Each site $i$ is initially assigned a stress $\sigma_i= \sigma_{\rm R}\pm 0.25 r$, where $r$ is a uniform random number between $\pm 1$. At each time step or plate update the initiating site, which is the site with the maximum stress, is found. The stress on all sites is increased by the same amount such that the stress on the initiating site equals $\sigma_F$. This procedure corresponds to the zero velocity limit of the loading plate in the Rundle-Jackson model~\cite{RJBModel}. When site $i$ fails, its stress is reset to the residual stress $\sigma_{R,i} = \sigma_{\rm R} + r \eta$ and the stress $(1-\alpha)[(\sigma_i-\sigma_{R,i})/4]$ is transferred to its four nearest neighbors. The magnitude of the noise is $\eta$. The value of the dissipation parameter, $\alpha$, is $0 \leq \alpha < 1$. Sites fail when the stress is greater than or equal to $\sigma_{\rm F}$. The failure of a site may cause neighboring sites to fail, triggering an avalanche. The stress is redistributed until the stress on all sites is less than $\sigma_{\rm F}$. This process concludes one plate update. 

Grassberger~\cite{grassberger} showed that the OFC model with periodic boundary conditions is deterministic for zero noise and that the dynamics appears to be stochastic for sufficiently high values of the noise. However, the transition between the two types of behavior was not explored.

In this paper, we show that there is a effective ergodicity breaking transition in the nearest-neighbor OFC model as the noise is varied. In Sec.~\ref{sec:break} we show that the OFC model is effectively ergodic~\cite{TM_1, TM_2, TM_3} for $\eta > \eta_c$, but is not effectively ergodic which implies not ergodic for $\eta <\eta_c$, where $\eta_c$ is the critical noise. In Sec.~\ref{sec:RP}, we characterize the high and low noise phases using recurrence plots, which are commonly used in nonlinear dynamics~\cite{RecMap}. From the recurrence plots of the stress on given sites, we calculate the recurrence rate~\cite{RecMap_Review} and define the recurrence fraction, $f_{\rm R}$, as the recurrence rate averaged over all sites. The recurrence fraction acts as a order parameter and differentiates the high and low noise phases.

We also use ideas from percolation theory to examine the critical behavior as the critical noise is approached from above. In Sec.~\ref{sec:CA} we treat all sites that fail in a plate update as part of the same cluster and determine the mean cluster size $\chi$ and the mean radius of gyration $R_{\rm G}$~\cite{stauffer1979scaling}. We determine the exponents $\gamma$ and $\nu$ associated with the divergence of $\chi$ and $R_G$ respectively as $\eta \to \eta_c$~\cite{stauffer1979scaling}. We also measure the Fisher exponents $\tau$ and $\sigma$ for the cluster distribution near the critical noise~\cite{fisher1967theory,stauffer1979scaling}. Our measured exponents are consistent with hyperscaling.

The numerical results reported in the following are for $\sigma_{\rm F} = 2.0$ and $\sigma_{\rm R} = 1.0$, $\alpha=0.01$, and $L=500$. Our results for $\alpha \in [0.005, 0.1]$ are qualitatively similar. For all our runs, we discarded at least the first $5 \times 10^{6}$ plate updates before recording data for $3 \times 10^{6}$ plate updates.

\section{\label{sec:break}Breakdown of Effective Ergodicity}
For systems with many degrees of freedom, it is difficult to verify if a system is ergodic and ergodicity can be checked rigorously for only a few simple systems~\cite{sinai1963foundations}. Instead, we use the stress fluctuation metric~\cite{TM_1,TM_2, TM_3} to determine if the system is effectively ergodic and to study the transition between phases that are not necessarily in equilibrium. The stress fluctuation metric describes how the spatial variance of the time-averaged stress on each site behaves for very long times. A spatially homogeneous system is effectively ergodic if the time average of the stress on each site approaches the same value. An analysis of the temporal properties of the metric can be found in Ref.~\cite{TM_2}.

\begin{figure}[t]
\includegraphics{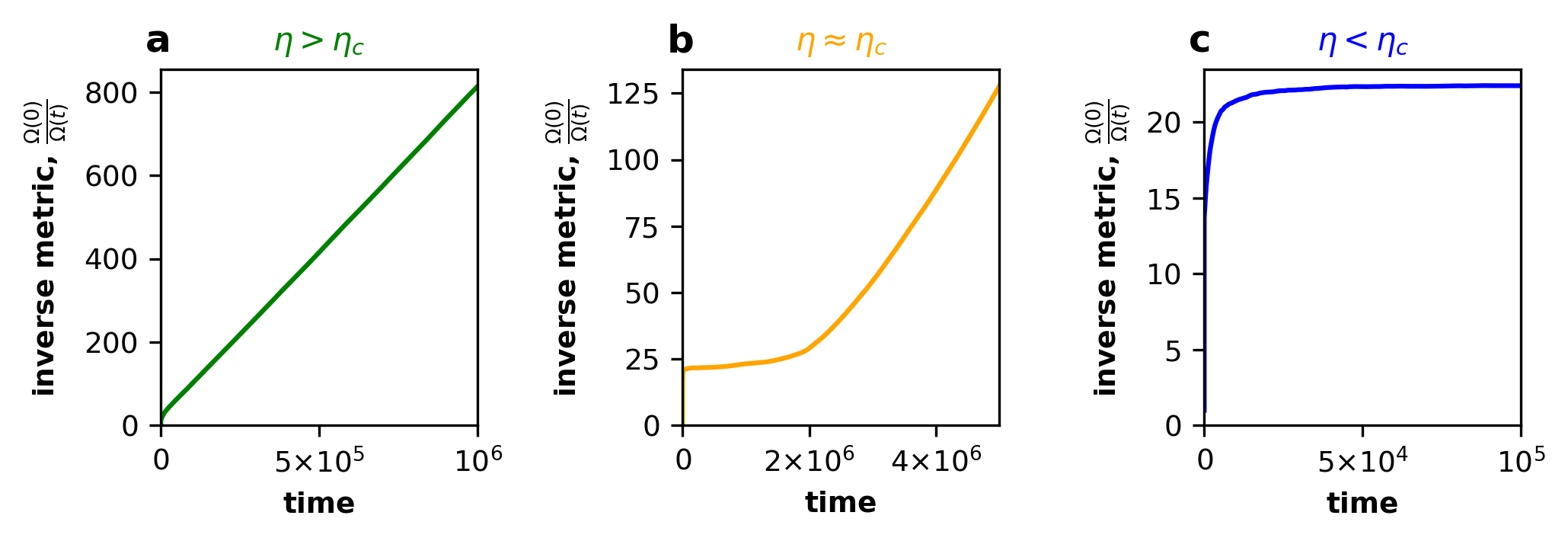}
\vspace{-0.2in}
\caption{The normalized inverse stress fluctuation metric as a function of time shows the breakdown of effective ergodicity at the critical noise. (a) For $\eta = 0.08> \eta_{c} $, $\Omega(0)/\Omega(t)$ is linear, and the system is effectively ergodic. (b) For $\eta = \eta_{c} \approx 0.071$, $\Omega(0)/\Omega(t)$ is initially flat, but becomes approximately linear during the observation time. (c) For $\eta = 0.06 < \eta_{c}$ the inverse metric is flat, indicating that the system is not ergodic.} 
\label{fig_metric}
\end{figure}

We define the time average of the stress at site $i$, up to time $t$, as
\begin{equation}
\overline{\sigma_i}(t) = \frac{1}{t} \sum_{t'=1}^t \sigma_{i}(t'),
\end{equation}
where $t$ represents the number of plate updates. The spatial average of the time-averaged stress on all sites is 
\begin{equation}
\lb \sigma(t) \rb = \frac{1}{N} \sum_{i=1}^N \overline{\sigma}_{i}(t).
\end{equation}
The stress fluctuation metric is defined as
\begin{equation}
\Omega(t) = \frac{1}{N}\sum_{i=1}^{N}[\overline{\sigma}_i(t) - \lb \sigma(t)\rb]^{2}.
\end{equation}
If the system is effectively ergodic, $\Omega(t)$ approaches zero as $1/t$~\cite{TM_1,TM_2, TM_3}. The system is not effectively ergodic during the observation time if the metric reaches a finite value, or does not increase linearly. The fluctuation metric provides a necessary but not sufficient condition for ergodicity.

As shown in Fig.~\ref{fig_metric}, the inverse metric increases linearly and the system is effectively ergodic for $\eta>\eta_{c} \approx 0.071$. For $\eta \approx \eta_{c}$, the inverse metric is initially flat for some time before showing a slow linear increase. For $\eta<\eta_{c}$, the inverse metric reaches a plateau, implying that the system is no longer effectively ergodic during our observation time. As the noise is increased past the critical noise, the system transitions from a phase that is not effectively ergodic during the observation time to one that is effectively ergodic. 

\begin{figure}[t]
\includegraphics[width=0.7\linewidth]{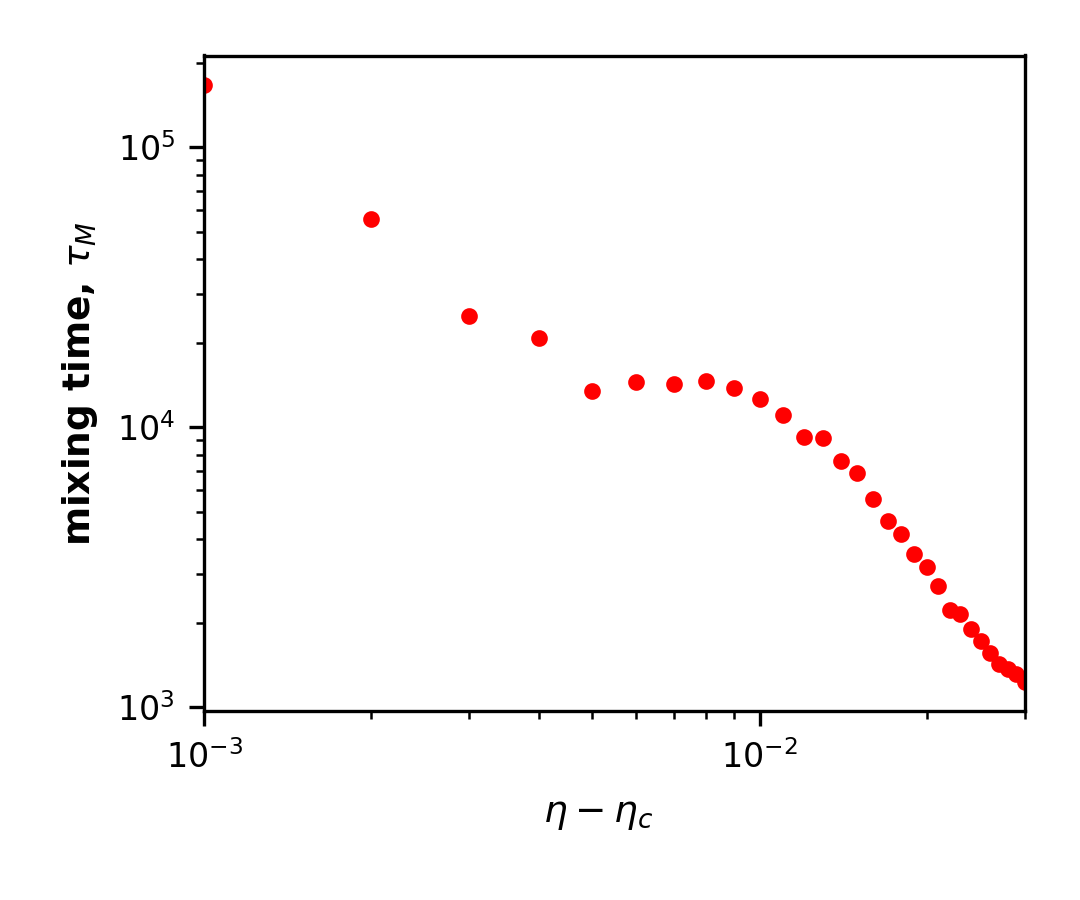}
\vspace{-0.2in}
\caption{ As the noise $\eta$ approaches the critical noise $\eta_c$ there is a rapid increase in the mixing time $\tau_{\rm M}$.} 
\label{fig_MixingTime}
\end{figure}

We can define the mixing time $\tau_M$ from the linear behavior of the inverse of the metric for $\eta > 
\eta_c$ as $\Omega(0)/\Omega(t) = t/\tau_{M}$. The mixing time is a measure of the how quickly the differences between the time-averaged stress on each site vanishes. Systems for which the differences never approach zero are not ergodic. The mixing time as a function of the noise $\eta$ is shown in Fig.~\ref{fig_MixingTime}. We see that there is an apparent divergence in $\tau_{M}$ as $\eta$ approaches $\eta_{c}$ from above, which implies the existence of critical slowing down.

\section{Recurrence Plots\label{sec:RP}}
We can analyze the dynamical transition using recurrence plots~\cite{RecMap_Review,RecMap} to explore the short term dynamics of the system. The recurrence plots are generated as follows. If the stress on a given site at times $t$ and $t'>t$ differs by less than $\epsilon$, then the corresponding point on the two-dimensional recurrence plot assumes the value one. Otherwise, the value is zero. From the time series of the stress on a given site, the recurrence matrix, $R(t,t')$, is defined as
\begin{equation}
R_{t,t'}(\epsilon) =
\begin{cases}
1 & \mbox{for } |\sigma(t')-\sigma(t)| < \epsilon 
\\
0 & \mbox{for } |\sigma(t')-\sigma(t)| \geq \epsilon.
\end{cases}
\end{equation}
A common choice for the threshold $\epsilon$ is 10\% of the range of values that the states can take~\cite{RecMap_Review}, which in our case is one, leading to the choice $\epsilon = 0.1$. We will show results only for $\epsilon=0.1$, but our results for other values of $\epsilon$ are consistent. 

\begin{figure}[t]
\includegraphics[width=1.0\linewidth]{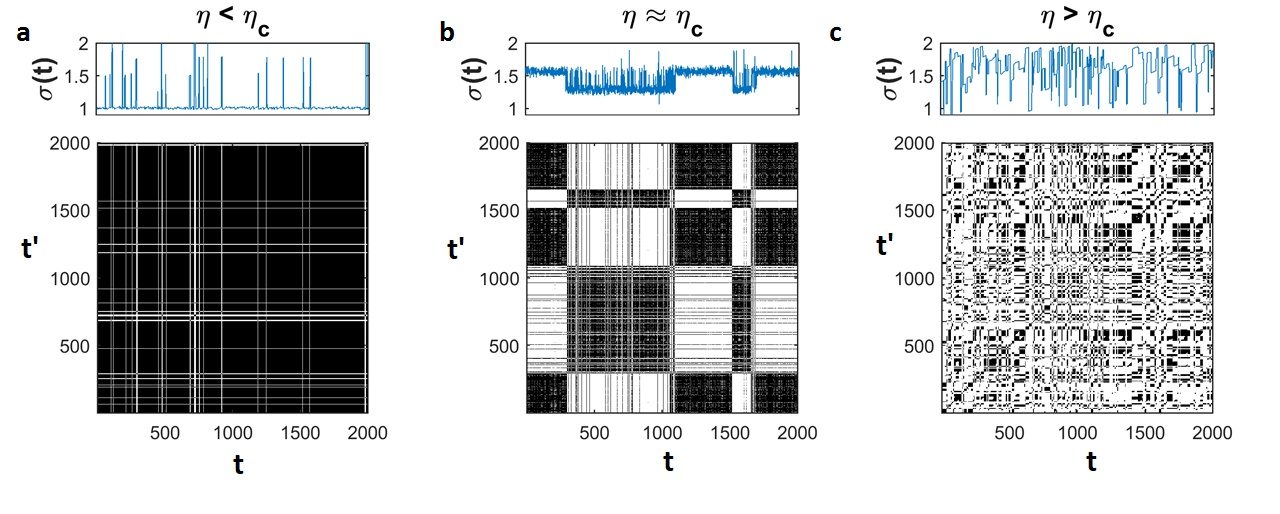}
\caption{The dynamics of the stress on a site transitions from recurrent to stochastic as the noise $\eta$ is increased past the critical noise $\eta_c$. The time series of the stress $\sigma(t)$ on a given site and the corresponding recurrence map is shown in the top and bottom rows, respectively. The dark points in the recurrence map corresponds to the recurrences. (a) For $\eta< \eta_{c}$ the dynamics is strongly recurrent, and the values of the stress on the site are confined to a small subset of the possible values. (b) For $\eta \approx \eta_{c}$, shown in the middle column, the dynamics is quasi-stationary. (c) If $\eta > \eta_{c}$, the dynamics is stochastic.} 
\label{fig_recurrence}
\end{figure}

The top row in Fig.~\ref{fig_recurrence} shows the time series of the stress on a given site and the bottom row shows the corresponding recurrence plot. For $\eta<\eta_{c}$ in Fig.~\ref{fig_recurrence}(a), the time series of the stress is confined to a narrow set of values. The recurrence plot confirms that the trajectory is strongly recurrent. For $\eta \approx \eta_{c}$ in Fig.~\ref{fig_recurrence}(b), the time series is quasi-stationary, and the recurrence map shows alternating black and white bands. The dynamics is stochastic for $\eta > \eta_{c}$, as shown by the time series and the corresponding recurrence map in Fig.~\ref{fig_recurrence}(c). We conclude from the recurrent plots that the dynamics of the stress on a given site undergoes a significant change as the noise is varied from $\eta < \eta_c$ to $\eta > \eta_c$. 

We use the recurrence plots to introduce a scalar order parameter to differentiate between the phases above and below the critical noise. The recurrence rate RR~\cite{RecMap_Review} measures the fraction of times the stress on a site returns to the neighborhood of its original value, averaged over all initial conditions. The recurrence rate RR for trajectories of duration, $T$, is given by
\begin{equation}
\mbox{RR}^{(i)}= \frac{1}{T} \sum_{t, t'=1}^{T} R^{(i)}_{t,t'} (\epsilon).
\label{eq_RR}
\end{equation}

\begin{figure}[t]
\includegraphics{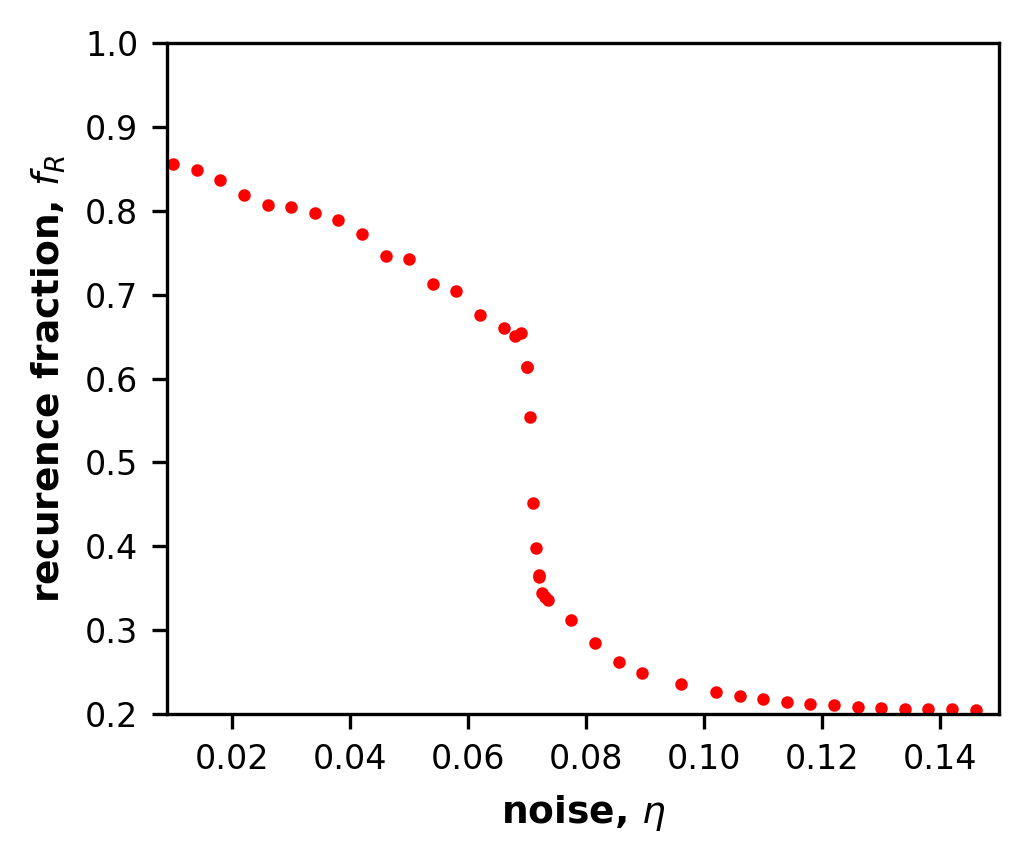}
\vspace{-0.2in}
\caption{The recurrence fraction $f_{\rm R}$ as a function of $\eta$ shows a transition from recurrent to stochastic dynamics for $\eta$ near $\eta_c$. }
\label{fig_recRate}
\end{figure}

We define the recurrence fraction $f_{\rm R}$ as the recurrence rate averaged over all sites. We find that $f_{\rm R}$ averaged over 50 sites, chosen at random, gives a good measure of the average recurrence rate. If $f_{\rm R}\approx 1$, the system is in the ordered state for which the trajectories of the stress on individual sites appear regular and are restricted to a subset of the allowed values of stress as shown in Fig.~\ref{fig_recurrence}(a). If $f_{\rm R}\approx 2\epsilon$, the system is in the disordered phase, and the local trajectories fluctuate almost randomly with no well defined order [see Fig.~\ref{fig_recurrence}(c)]. In Fig.~\ref{fig_recurrence}(b), we see quasi-periodic changes in the stress trajectory, and $f_{R}$ for $\eta \approx \eta_c$ is larger than for $\eta<\eta_c$. Note that $f_{\rm R}$ in Fig.~\ref{fig_recRate} appears to show a discontinuous jump at $\eta_{c}$. However, we are limited by the rapid increase of the mixing time near $\eta \approx \eta_c$ in determining if there is an actual jump in the order parameter. The rapid change in $f_R$ as the noise is varied implies a divergence in the fluctuations, which we will explore using the cluster analysis of the failed sites. 

\section{Cluster Analysis\label{sec:CA}}
\begin{figure}[h]
\includegraphics[width=0.75\linewidth]{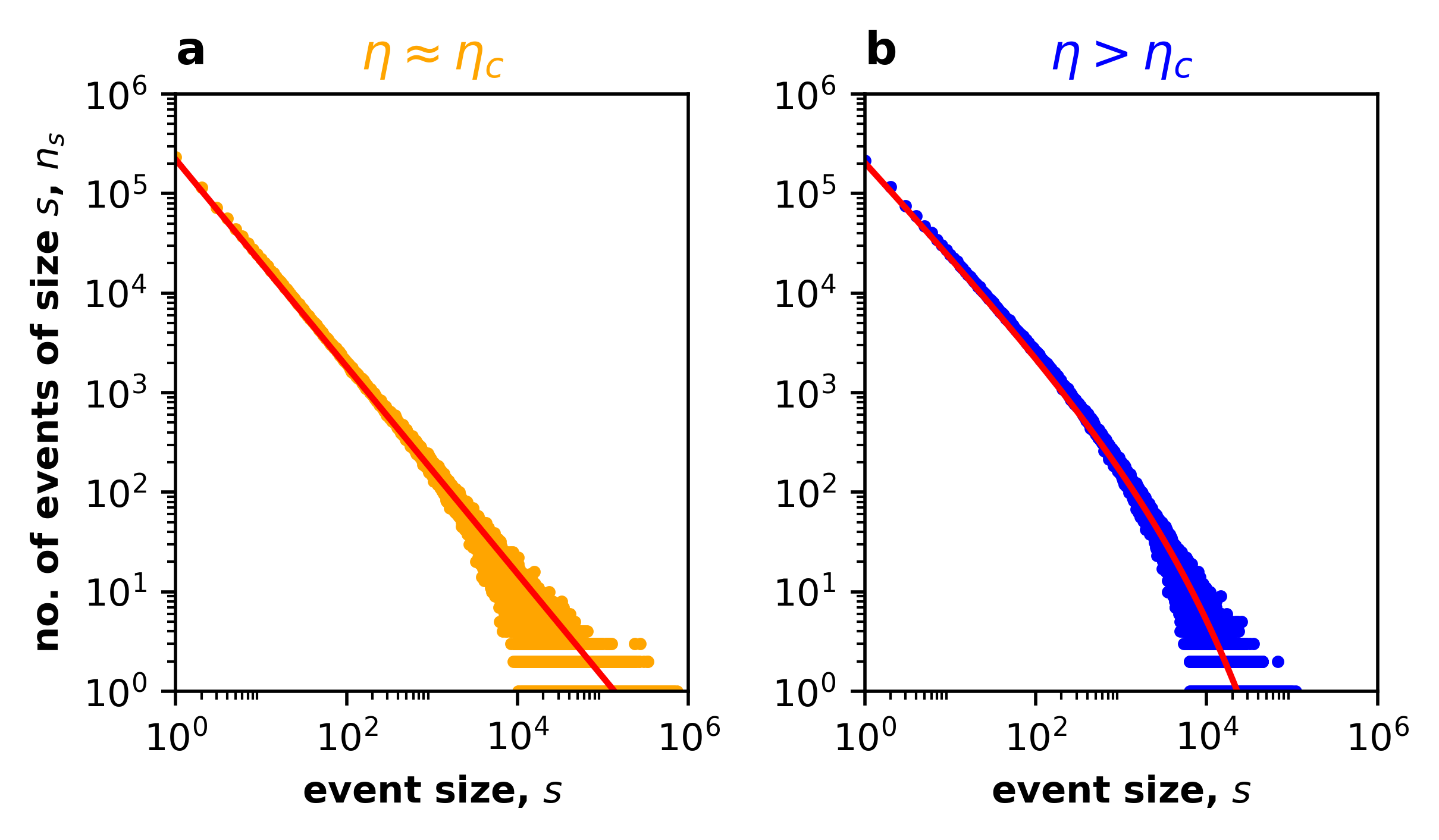}
\caption{The measured value of Fisher exponents $\tau$ and $\sigma$ is consistent with the hyperscaling laws. The number of clusters of size $s$, $n_s$, can be fitted to Eq.~\eqref{eq_fisher} to estimate the exponents $\tau$ and $\sigma$~\cite{fitPython}. (a) We find $\tau \approx 1.04 \pm 0.14$ for $\eta=0.071$. (b) For $\eta = 0.08$ we estimate $\sigma=0.43\pm 0.03$. } 
\label{fig_scal}
\end{figure}

In equilibrium statistical mechanics, we can learn about the nature of a transition from the geometric properties of the fluctuations~\cite{coniglio1980clusters, klein_big} for certain systems. We map the failed sites onto a percolation problem by assuming that an event of $s$ sites corresponds to a percolation cluster~\cite{Klein_GR, coniglio1980clusters, Serino_stress, fisher1967theory}. When the noise is greater than $\eta_c$, the system is effectively ergodic and the distribution of the clusters can be fit to a power law with an exponential cutoff as
\begin{equation}
n_{s} \sim {s^{-\tau} \exp(-(\eta-\eta_c) s^{\sigma})} ,
\label{eq_fisher}
\end{equation}
where $n_{s}$ is the number of clusters with $s$ failed sites and $\tau$ and $\sigma$ are the Fisher exponents. From Fig.~\ref{fig_scal} these exponents are estimated to be $\tau \approx 1.04 \pm 0.14$, and $\sigma=0.43\pm 0.03$.

\begin{figure}[t]
\includegraphics{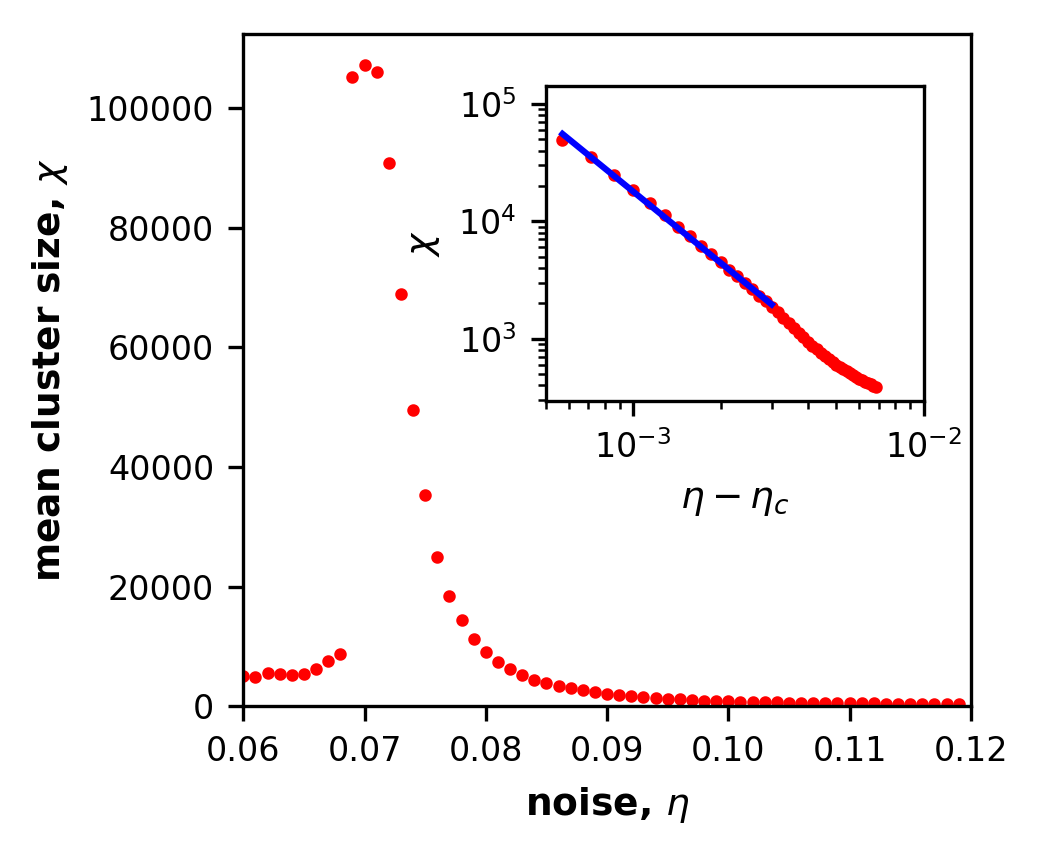}
\vspace{-0.2in}
\caption{The critical exponent $\gamma$ determined from the power law behavior of the mean cluster size, $\chi$, is estimated to be $\approx 2.01 \pm 0.14$. } 
\label{fig_meanExp}
\end{figure}

\begin{figure}[t]
\includegraphics{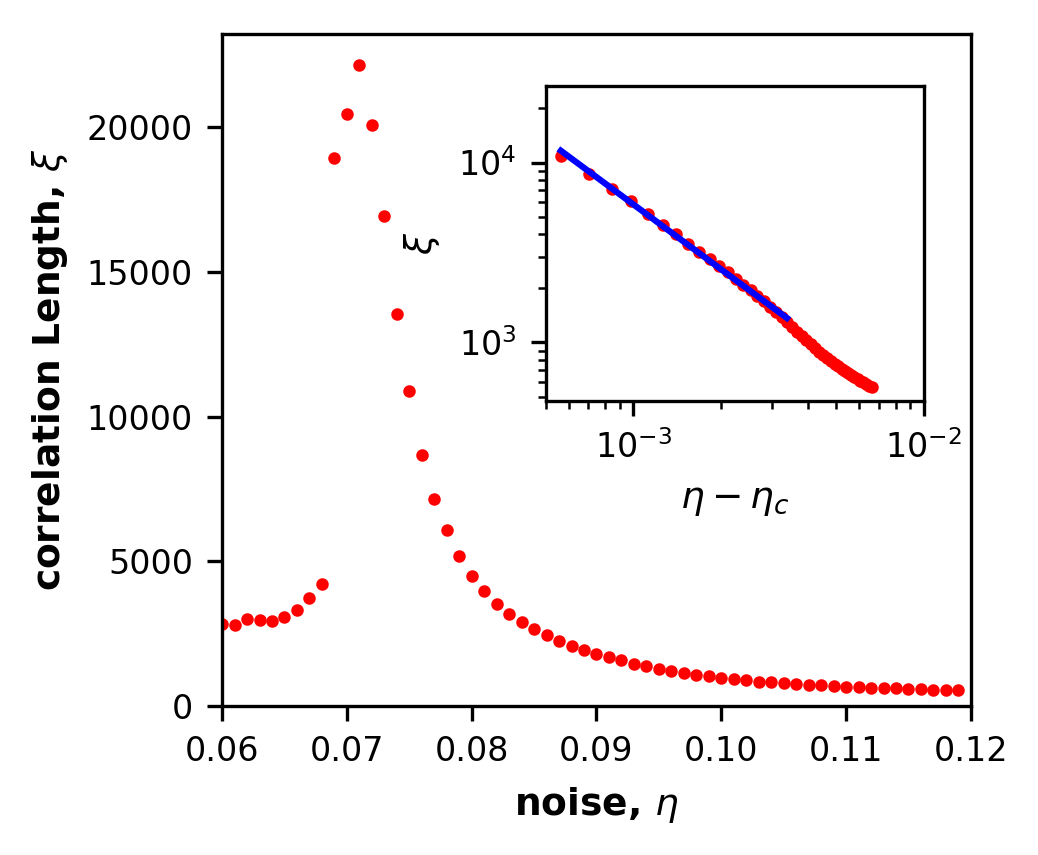}
\vspace{-0.2in}
\caption{The critical exponent $\nu$ characterizes the divergence of the connectedness length,$\xi$, and is estimated to be $\approx 1.20 \pm 0.13$. } 
\label{fig_corrExp}
\end{figure}

We measure the mean event size (analogous to the susceptibility in thermal systems) $\chi$ and the connectedness length (analogous to the correlation length) $\xi$, which are defined as~\cite{stauffer1979scaling}
\begin{align}
\chi & = \frac{\sum_{s} s^{2} n_s}{\sum_{s} s n_s} \\
\noalign{\noindent and}
\xi & = \frac{\sum_{s} s^{2}{R^{2}_{\rm G}(s)} n_s }{\sum_{s} s n_s},
\end{align}
where, $R_{\rm G}(s)$ is the mean radius of gyration of clusters with $s$ sites~\cite{stauffer1979scaling}. From Fig.~\ref{fig_meanExp} we see that $\chi$ diverges as $(\eta-\eta_c)^{-\gamma}$ with the exponent $\gamma \approx 2.01\pm 0.14$. Similarly, from Fig.~\ref{fig_corrExp} we find that $\xi \sim (\eta- \eta_c)^{-\nu}$ with $\nu \approx 1.25 \pm 0.13$. The divergence of the mean cluster size is used to estimate that the critical noise is $\eta_{c}\approx0.071$. This value of the critical noise is consistent the value of $\eta$ at which the jump in the recurrence fraction occurs as seen in Fig~\ref{fig_recRate}. We are unable to determine the heat capacity and the exponent $\alpha$ directly because we have not identified an energy in the nearest-neighbor OFC model.

Although the exponents $\gamma$, $\nu$, $\tau$, and $\sigma$ were estimated independently, they are related by hyperscaling~\cite{hyper}. To test the consistency of the measured exponents with the scaling relations, we need to add one to the measured value of $\tau$ from our simulations, because the clusters in the OFC model are grown from a seed site~\cite{stauffer1979scaling, Serino_stress}. We define $\tilde{\tau}=\tau +1$ as the corrected exponent and write hyperscaling laws are $\gamma = (3 - \tilde{\tau})/\sigma$ and $\nu = (\tilde{\tau} - 1)/d \sigma$, where $d$ is the spatial dimension.  If we substitute $\gamma = 2$, $\nu=5/4$, and $d=2$, we find the Fisher exponents to be $\tau=10/9$ and $\sigma=4/9$. Our numerical estimates of $\tau \approx 1.04 \pm 0.14$ and $\sigma=0.43\pm 0.03$ are consistent with the predictions from hyperscaling. If we use the additional scaling relations, $\alpha = 2 - (\tilde{\tau}-1)/\sigma$ and $\beta = (\tilde{\tau} - 2)/\sigma$, we find $\beta=1/4$ and $\alpha=-1/2$, corresponding to the critical behavior of the order parameter and the heat capacity. We are unable to estimate $\beta$ directly using $f_{\rm R}$ because $\beta$ is close to zero and the range of $\eta$ over which power law behavior is observed is too small.

\begin{figure}[t]
\includegraphics[width=0.85\linewidth]{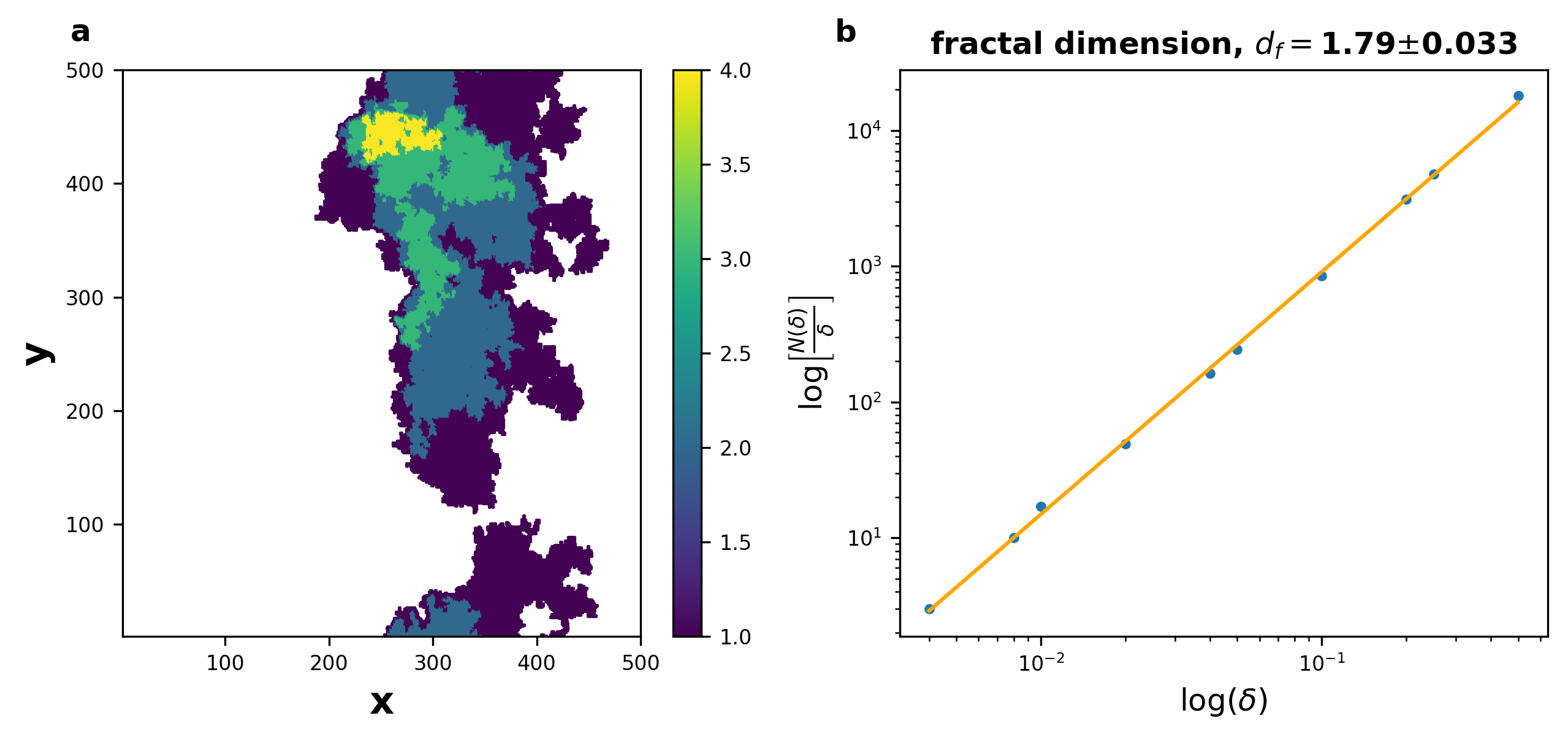}
\vspace{-0.2in}
\caption{ The measured value of the fractal dimension $d_f$ is consistent with predictions of hyperscaling. (a) The cluster of failed sites appears to be a fractal. (b) The fractal dimension was computed using a box-counting method. The number of boxes needed to cover the cluster is $N(\delta)$, where $\delta$ is the box length. In the limit, that the box length $\delta \to 0$, the slope of $\ln\left[N(\delta)/\delta\right]$ corresponds to the fractal dimension. The estimated fractal dimension is $d_f=1.79\pm 0.03$. } 
\label{fig_df}
\end{figure}

Another exponent that we can measure independently is the fractal dimension, $d_f$, of the clusters. Using the scaling relations we can express the fractal dimension in terms of the measured exponents and the spatial dimension; $d_f = d - \beta/\nu = d - (\tilde{\tau}-2)/(\sigma \nu)$. Our measured value of $d_f=1.79\pm0.03$ is consistent with the prediction from hyperscaling.

\section{Long-Rnge stress transfer \label{sec:LR}}
\begin{figure}[t]
\includegraphics{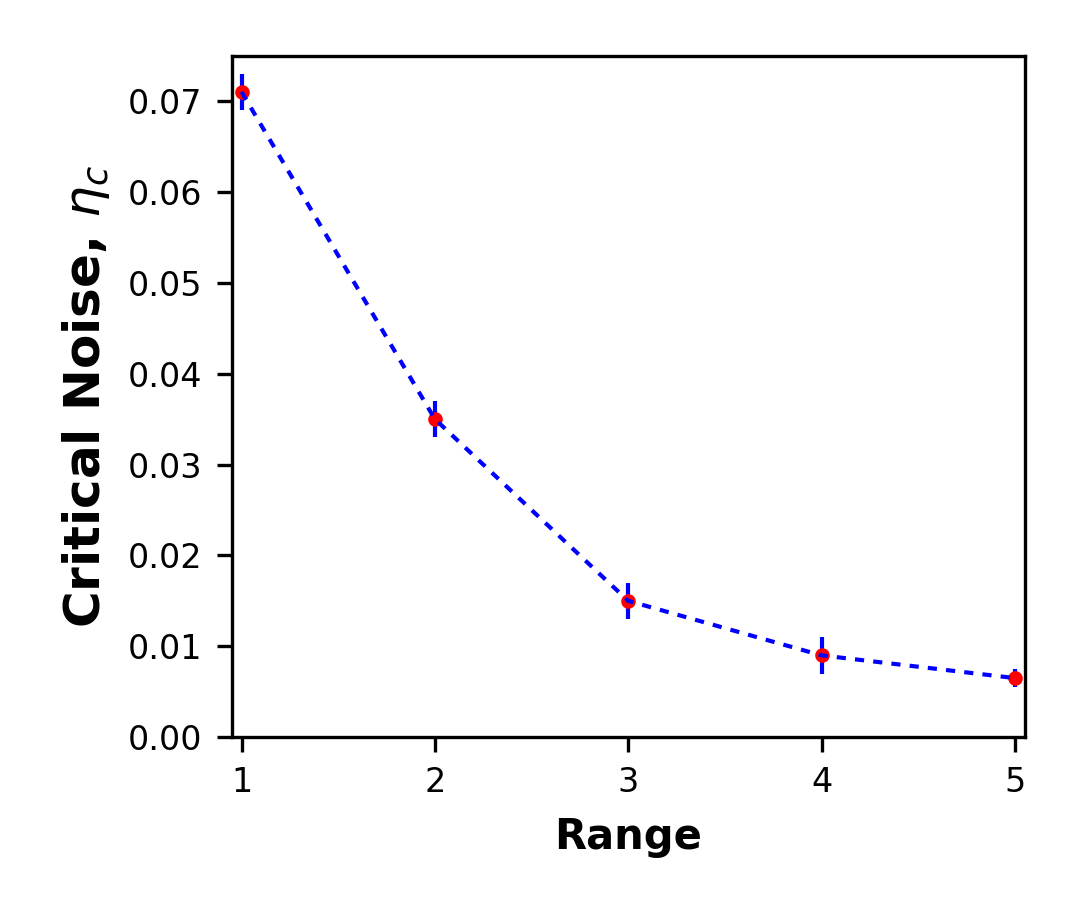}
\vspace{-0.2in}
\caption{ The critical noise, $\eta_c$, in the OFC model decreases as the range of stress transfer is increased. } 
\label{fig_R-Noise}
\end{figure}

Our results for the nearest neighbor OFC model also apply to long-range stress transfer. In this case, a failing site distributes its stress equally to all sites within a circle of radius, $R$, the stress transfer range. The nearest neighbor case corresponds to the $R=1$. Figure~\ref{fig_R-Noise} shows that the value of the critical noise decreases as the range of stress transfer increases. For each value of the range $R$, the jump in the recurrence fraction $f_{\rm R}$, the divergence in the mixing time $\tau_{\rm M}$ and the mean cluster size, $\chi$, occur at the same value of noise. 

\section{Discussion \label{sec:Disc}}
We have found a novel phase transition in the OFC model as the noise is varied for all stress transfer ranges studied. Below the critical noise $\eta_{c}$ , the system is not ergodic and the stress on the sites appears to evolve via limit cycles. In contrast, for $\eta>\eta_c$ the system is effectively ergodic and the dynamics is stochastic . For $\eta<\eta_c$ recurrence plots show that individual sites are trapped in limit cycles with long lifetimes. As $\eta$ approaches $\eta_c^-$, the limit cycles become unstable, and the trajectories of individual sites show deviations from  limit cycle behavior. Larger values of $\eta$ disrupt the limit cycles and for $\eta>\eta_c$ the trajectories appear random. We find that the recurrence fraction, $f_R$, is a convenient choice of the order parameter and describes the transition from limit cycle to stochastic behavior. The $f_R$ appears to exhibit a discontinuous jump at $\eta = \eta_c$. 

We also investigated the transition for $\eta\to \eta_c^{+}$ using a cluster analysis to study properties such as the mean cluster size, the connectedness length, and the Fisher exponents. Our measured numerical values of the exponents are consistent with hyperscaling within statistical error. An unusual feature of this ``percolation'' description of the transition is that there is no ``infinite cluster'' or avalanche for $\eta$  above or below $\eta_{c}$. Hence the percolation exponent $\beta$ associated with how the probability that a site selected at random belongs to the infinite cluster goes to zero as the transition is approached cannot be measured directly. However, a measurement of the fractal dimension $d_{f} = d - \beta/\nu$ yields a value of $\beta$ consistent with the relation $\gamma + 2\beta = d\nu$. 

Our results indicate that the effect of noise in non-equilibrium phase transitions is more subtle than was previously suspected and that noise may play a significant role in transitions in systems such as Kuramoto model~\cite{Review_Kuramoto} and neural systems modeled by integrate-and-fire neurons~\cite{neuron}. Of particular interest is the role of noise in the behavior of earthquake faults. The OFC model(and the virtually identical Rundle-Jackson model) with long-range stress transfer has been used to study earthquake fault systems~\cite{Klein_GR, Serino_stress}, where the stress transfer occurs over large distances due to elastic forces. Figure~\ref{fig_R-Noise} shows that the critical noise $\eta_c$ decreases as the stress transfer range is increased. The noise in real earthquake fault systems is believed to be small~\cite{rock-st}, so it is possible that earthquake faults operate near the critical noise. Therefore, small changes in the noise due to variations in water content in rocks or microcrack density can drive the fault system into a different phase where the dynamics are considerably different. Our result suggests one possible mechanism by which a fault may change its behavior from quasi-periodic to scale-free distribution of events consistent with Gutenberg-Richter scaling~\cite{wesnousky1994gutenberg}.

\begin{acknowledgements}
We would like to thank Tyler Xuan Gu for prompting the study of the OFC model for low noise and Erik Lascaris, Rashi Verma, and Shan Huang for helpful comments. 
\end{acknowledgements}


\begin{thebibliography}{99}
\bibitem{fitPython}
Python's scipy.optimize\_curvefit was used for the nonlinear fits.
\end{thebibliography}
\end{document}